\documentclass{jfm}

\newcommand*{\rom}[1]{\expandafter\@slowromancap\romannumeral #1@}
\makeatother
\usepackage{amsmath,amssymb}
\usepackage{lipsum}  
\usepackage{graphicx}
\usepackage{dcolumn}
\usepackage{bm}
\usepackage{subcaption}
\usepackage{comment}
\usepackage{xcolor}
\usepackage{lipsum}
\definecolor{red}{rgb}{1, 0, 0}
\definecolor{blue}{rgb}{0, 0, 1}
\definecolor{green}{rgb}{0, 0.5, 0}
\usepackage{etoolbox}
\usepackage{tikz}
\usepackage{placeins} 
\usepackage[normalem]{ulem}
\newcommand{\pd}{\partial}
\newcommand{\avg}[1]{\langle{#1}\rangle}

\newcommand{\what}{\widehat}
\newcommand{\tog}[1]{\mbox{\(#1\)}} 

\newcommand{\Ra}{\mathit{Ra}}
\renewcommand{\Pr}{\mathit{Pr}}
\newcommand{\Nu}{\mathit{Nu}}
\newcommand{\Rh}{\mathit{Rh}}

\shorttitle{Two-dimensional Rayleigh-B\'enard convection without boundaries}
\shortauthor{P.~Winchester, V.~Dallas and P.~D.~Howell}

\title{Two-dimensional Rayleigh-B\'enard convection without boundaries}

\author{
Philip Winchester\aff{1}
  \corresp{\email{winchester@maths.ox.ac.uk}},
   Vassilios Dallas \aff{1}
  \corresp{\email{dallas@maths.ox.ac.uk}},
 \and Peter~D.~Howell\aff{1}
 \corresp{\email{howell@maths.ox.ac.uk}}
  }

\affiliation{\aff{1}Mathematical Institute, University of Oxford, Oxford, OX2 6GG, UK
}

\begin{document}

\maketitle

\begin{abstract}
We study the effects of Prandtl number $\Pr$ and Rayleigh number $\Ra$ in two-dimensional Rayleigh-B\'enard convection without boundaries, i.e. with periodic boundary conditions. In the limits of $\Pr \to 0$ and $\infty$, we find that the dynamics are dominated by vertically oriented elevator modes that grow without bound, even at high Rayleigh numbers and with large scale dissipation. For finite Prandtl number in the range $10^{-3} \leq \Pr \leq 10^2$, the Nusselt number tends to follow the `ultimate' scaling $\Nu \propto \Pr^{1/2}\Ra^{1/2}$, and the viscous dissipation scales as $\epsilon_\nu \propto \Pr^{1/2}\Ra^{-1/4}$. The latter scaling is based on the observation that enstrophy $\avg{\omega^2} \propto \Pr^0 \Ra^{1/4}$. The inverse cascade of kinetic energy forms the power-law spectrum $\what E_u(k) \propto k^{-2.3}$, while the direct cascade of potential energy forms the power-law spectrum $\what E_\theta(k) \propto k^{-1.2}$, with the exponents and the turbulent convective dynamics in the inertial range found to be independent of Prandtl number. Finally, the kinetic and potential energy fluxes are not constant in the inertial range, invalidating one of the assumptions underlying Bolgiano-Obukhov phenomenology.
\end{abstract}

\section{Introduction} \label{sec:Intro}

The fundamental challenge to understand thermally driven flow in the strongly nonlinear regime has puzzled the fluid dynamics community for more than a century \citep{Doering19}. The typical set-up that is studied theoretically consists of a fluid confined between two horizontal plates that is heated from below and cooled at the top. This is the so called Rayleigh-B\'enard convection (RBC) problem after Rayleigh's proposed model \citep{rayleigh1916} for B\'enard's experiment on buoyancy-driven thermal convection \citep{benard1900,benard1927}.
The classical problem depends on two dimensionless parameters: the Rayleigh number $\Ra$, which measures the driving effect of buoyancy relative to the stabilising effects of viscosity and thermal diffusivity; and the Prandtl number $\Pr$, which is the ratio of kinematic viscosity to thermal diffusivity.
The global flow properties may be characterised by the Nusselt number $\Nu$, a further dimensionless parameter that measures the total heat flux relative to the purely conductive heat flux.

For stably stratified turbulence, \cite{Bolgiano59} and \cite{Obukhov59} proposed the so-called Bolgiano--Obukhov (BO) phenomenology, in which buoyancy balances inertia in the momentum equation and the potential energy flux is approximately constant for length scales larger than the Bolgiano scale. These assumptions lead to
kinetic and potential energy spectra of the form
$\what E_u(k) \propto 
k^{-11/5}$
and
$\what E_\theta(k) \propto 
k^{-7/5}$,
respectively, where $k$ is the wavenumber. 
Despite originally being proposed for stably stratified flows, BO scaling has been reported for three-dimensional (3D) convection \citep{wuetal90,Calzavarini02,MishraVerma10,KaczorowskiXia13}. However, its existence is still debatable \citep{LohseXia10,KunnenClercx14}, with some studies reporting \cite{Kolmogorov41} power-law scaling $\propto k^{-5/3}$ for the kinetic and potential energy spectra \citep{BorueOrszag97,Verma17,VermaBook18}.
Similar debate remains in simulations of two-dimensional (2D) RBC with periodic boundary conditions, with some studies to argue for  \citep{Mazzino17,Toh94,xiehuang22} and some against \citep{BiskampSchwarz97,biskampetal01,celanietal01,celanietal02} the validity of BO scaling.

For the strongly nonlinear regime of thermal convection, which is of paramount importance for geophysical and astrophysical applications, there are two competing theories for the behaviour of 
$\Nu$ as $\Ra$ tends to infinity for arbitrary $\Pr$. 
These two proposed asymptotic scaling laws are 
the `classical' theory $\Nu \propto \Pr^0\Ra^{1/3}$ by \cite{malkus1954} and the `ultimate' theory $\Nu \propto \Pr^{1/2}\Ra^{1/2}$ by \cite{Kraichnan62, Spiegel71}. 
The Rayleigh number at which the transition to the ultimate scaling is presumed to occur is not known, and laboratory experiments and numerical simulations \citep{Niemela00,Chavanne01,Niemela03,Urban2011,He12,Zhuetal18,Doeringetal19} have reported different power-laws for a wide range of Rayleigh numbers.


Boundary conditions and boundary layers strongly affect the turbulence properties of RBC \citep{stellmachetal14,lepotetal18}. 
In the analogy of RBC with geophysical phenomena, the top and bottom boundaries are often absent particularly when the focus is to understand the dynamics of the bulk flow. In this study, we choose the most obvious theoretical approach to bypass boundary layer effects by considering a fully periodic domain \citep{Toh94,BorueOrszag97,biskampetal01,celanietal02,ngetal18} for the Rayleigh-B\'enard problem, with an imposed constant, vertical temperature gradient. For this homogeneous RBC set-up has been claimed that $Nu \propto Ra^{1/2}$ \citep{LohseToschi03,calzavarinietal05} in line with the ultimate theory and this scaling is also suggested from simulations of axially periodic RBC in a vertical cylinder \citep{schmidtetal12}.

Homogeneous RBC, however, exhibits exponentially growing solutions in the form of axially uniform vertical jets, called ``elevator modes'' \citep{BainesGill69,BatchelorNitsche91,calzavarinietal06,LiuZikanov15}. As soon as these modes grow to a significant amplitude, they experience secondary instabilities ultimately leading to statistically stationary solutions \citep{calzavarinietal06,schmidtetal12}. In recent numerical simulations the elevator modes were suppressed by introducing an artificial horizontal buoyancy field \citep{xiehuang22} or large-scale friction \citep{BarralDubrulle23}. The inverse cascade that is observed in 2D homogeneous RBC \citep{Toh94,VermaBook18,xiehuang22} is another source of energy to the large-scale modes that grow to extreme values forming a condensate, whose amplitude saturates when the viscous dissipation at the largest scale balances the energy injection \citep{Chertkovetal2007,BoffettaEcke2012}. So, in this study to avoid the unbounded growth of energy we include large scale dissipation to mimic the effect of friction when there are boundaries and to be able to reach statistically stationary solutions. 


Most of the attention on two-dimensional (2D) RBC focuses is on the Rayleigh number dependence of the dynamics, with only a few studies (e.g., \citep{calzavarinietal05,zhangetal17,heetal21}) considering the effects of the Prandtl number. 
In this paper, we extensively study the effects of the Prandtl number and the Rayleigh number
using numerical simulations of 2D RBC in a periodic domain driven by constant temperature gradient, while also considering hyperviscous simulations to permit the large scale separation, which is crucial for the analysis of the multi-scale dynamics.
Sec.~\ref{sec:ProblemDescription} contains the dynamical equations, numerical methods and the definitions of the spectral and global variables under study. 
The results of our simulations are presented in Sec.~\ref{sec:results}.
After briefly investigating
the behaviour of the system in the limits of zero and infinite Prandtl number,
we analyse how the global variables scale with Prandtl and Rayleigh numbers and then discuss the effects of these dimensionless parameters on the spectral dynamics. Finally, we summarise our conclusions in Sec.~\ref{sec:Conclusions}.

\section{Problem Description} \label{sec:ProblemDescription}

\subsection{Governing equations}

We consider two-dimensional Rayleigh–B\'enard convection 
of a fluid heated from below in a periodic square cell $(x,y)\in[0,L]^2$.
The temperature $T(x,y,t)$ is decomposed as $T=-\Delta T y/L+\theta$, where $\Delta T/L$ is the constant imposed temperature gradient and the temperature perturbation $\theta(x,y,t)$ satisfies periodic boundary conditions.
As usual, for simplicity we employ the Oberbeck--Boussinesq approximation \citep{oberbeck1879,boussinesq1903,tritton12}, in which the kinematic viscosity $\nu$ and the thermal diffusivity $\kappa$ are taken to be constant
while temperature dependence of the fluid density $\rho$ is neglected
except in the buoyancy term of the momentum equation. 

The governing equations of the problem in two dimensions can be written in terms of
$\theta(x,y,t)$ and the
streamfunction $\psi(x, y, t)$
as follows:
\begin{subequations}
\label{eq:gov_eq}
\begin{align}
    \label{eq:psi}
    \partial_t \nabla^{2} \psi + \{\psi, \nabla^2 \psi\} &=  \alpha g \pd_x \theta + (-1)^{n+1}\nu \nabla^{2n+2} \psi + \mu \psi, \\
    \label{eq:theta}
    \partial_t \theta + \{\psi, \theta \} &= \frac{\Delta T}{L} \pd_x \psi + (-1)^{n+1}\kappa \nabla^{2n} \theta,
\end{align}
\end{subequations}
where $\{A, B\} = \pd_x A \pd_y B - \pd_y A \pd_x B$ is the standard Poisson bracket, $\alpha$ is the thermal expansion coefficient and $g$ is the gravitational acceleration. 
To prevent the formation of a large scale condensate in the presence of an inverse cascade, to avoid the elevator modes and to reach a turbulent stationary regime we supplement our system with a large scale dissipative term $\mu \psi$ that is responsible for saturating the inverse cascade.
We consider both normal and hyper viscosity by raising the Laplacian to the power of $n=1$ and $n=4$, respectively. The hyperviscous case, albeit not physically realisable, gives a wider inertial range, as diffusive and viscous terms kick in abruptly at much smaller scales compared to the normal viscosity case.
In the limit of $\nu \to 0$, $\kappa \to 0$, and $\mu \to 0$ the quantity that is conserved is $E_u - \frac{agL}{\Delta T} E_\theta$, where the kinetic energy $E_u$ and the potential energy $E_\theta$ are defined by
\begin{equation}
E_u =\frac{1}{2}\avg{|\nabla\psi|^2},
\qquad
E_\theta = \frac{1}{2}\avg{\theta^2},
\end{equation}
with the angle brackets $\avg{\cdot}$ here denoting the spatiotemporal average.

Equations \eqref{eq:gov_eq} depend on three dimensionless parameters, namely
\begin{equation}
\Pr = \frac{\nu}{\kappa}, \qquad 
\Ra = \frac{\alpha g\Delta T L^{4n-1}}{\nu \kappa}, \qquad 
\Rh = \mu \left(\frac{L^5}{\alpha g \Delta T}\right)^{1/2},
\label{eq:params}
\end{equation}
which are the Prandtl number, Rayleigh number and friction Reynolds number, respectively,
in accordance with
\begin{equation}
    \label{eq:scaling}
    {\bf x} \sim L, \quad t \sim \frac{L^{2n}}{\kappa}, \quad \psi \sim \frac{\kappa}{L^{2n-2}}, \quad \theta \sim \Delta T.
\end{equation}

\begin{figure}
\centering
 \includegraphics[width=0.7\linewidth]{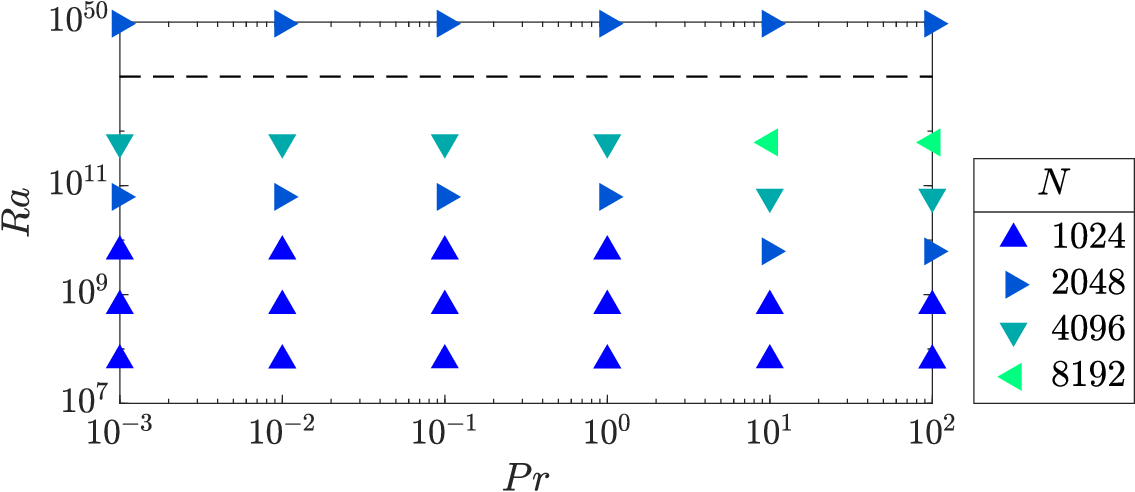}
 \caption{\label{fig:ParamSpace} 
    Parameter values simulated in the $(\Ra,\Pr)$-plane with the resolution, $N$, used at each instant colour-coded in the legend. The black dashed line separates runs with normal viscosity (below line) from those with hyperviscosity (above line). 
   }
\end{figure}

We perform direct numerical simulations (DNS) 
of Eqs.~\eqref{eq:gov_eq} using the pseudospectral method \citep{Orszag77}. We decompose the stream function into basis functions with Fourier modes in both the $x$ and $y$ directions, viz.
\begin{align}
    \label{eq:Basis}
    \psi({\bf x},t) &= \sum^{N/2}_{{\bf k} = -N/2} \widehat{ \psi}_{\bf k}(t) e^{i {\bf k} \cdot \mathbf{x}},
\end{align}
where $\widehat{ \psi}_{\bf k}$ is the amplitude of the ${\bf k} = (k_x, k_y)$ mode of $\psi$, and $N$ denotes the number of aliased modes in the $x$- and $y$-directions. We decompose $\theta$ in the same way. A third-order Runge-Kutta scheme is used for time advancement and the aliasing errors are removed with the two-thirds dealiasing rule \citep{mpicode05b}. 
In both the normal and hyperviscous simulations,
we find that $\Rh=(2\pi)^{5/2}$ yields
a saturated turbulent state that dissipates enough kinetic energy at large scales such that the kinetic energy spectrum peaks at $k = 2$ without over-damping the system.
So, we fix $\Rh = (2\pi)^{5/2} \simeq 100$ while varying the Rayleigh and Prandtl numbers in the ranges $6.2 \times 10^7 \leq \Ra \leq 6.2 \times 10^{11}$ and $10^{-3} \leq \Pr \leq 10^2$. To model large Rayleigh number dynamics, we set $\Ra = 9.4 \times 10^{49}$ in our hyperviscous simulations. Fig.~\ref{fig:ParamSpace} shows the parameter values simulated in the $(\Ra,\Pr)$-plane as well as the resolution, $N$, used in each case. 
Time-averaged quantities are computed 
over 1000 realisations once the system has reached a statistically stationary regime, sufficiently separated in time (at least 5000 numerical time steps) to ensure statistically independent realisations.
 
\subsection{Global and spectral quantities}

Next we briefly outline the global and spectral flow properties that will be explored in our numerical simulations below.
The energy spectra of the velocity field $\what E_u(k,t)$ and the temperature field $\what E_{\theta}(k,t)$, referred to as the kinetic energy and potential energy spectra,
are defined as
\begin{subequations}
\label{eq:Spectra}
\begin{align}
    \label{eq:KESpectra}
    \what E_u(k,t) &= \frac{1}{2}\sum_{k \leq |{\bf k}| < k+\Delta k} k^2\left| \widehat \psi_{\bf k}(t) \right|^2, \\
    \label{eq:PESpectra}
    \what E_{\theta}(k,t) &= \frac{1}{2}\sum_{k \leq |{\bf k}| < k+\Delta k} \left| \widehat \theta_{\bf k}(t) \right|^2,
\end{align}
\end{subequations}
where the sum is performed over the Fourier modes with wavenumber  
amplitude $k = |{\bf k}| = \sqrt{k_x^2 + k_y^2}$ in a shell of width $\Delta k = 2\pi/L$. 
Using the Fourier transform, one can derive the evolution equations of kinetic and potential energy spectra from Eqs.~\eqref{eq:gov_eq}, namely
\begin{subequations}
\label{eq:SpecEvolution}
\begin{align}
    \label{eq:KESpecEvolution}
    \partial_t \what E_u(k,t) &=  - \partial_k \Pi_u(k,t) - D_\nu(k,t) - D_\mu(k,t) + \alpha g F_B(k,t), \\
    \label{eq:PESpecEvolution}
    \partial_t \what E_{\theta}(k,t) &= -\partial_k \Pi_\theta(k,t) - D_\kappa(k,t) + \frac{\Delta T}{L}F_B(k,t).
\end{align}
\end{subequations}

The energy flux $\Pi$ is a measure of the nonlinear cascades in turbulence \citep{Alexakis18}. The energy flux for a circle of radius $k$ in the 2D wavenumber space is the total energy transferred from the modes within the circle to the modes outside the circle. Consequently, we define the flux of kinetic energy $\Pi_u(k,t)$ and potential energy $\Pi_{\theta}(k,t)$ as 
\begin{subequations}
\label{eq:Flux}
\begin{align}
    \label{eq:KEFlux}
    \Pi_u(k,t) &= \sum_{k' \leq k} T_u(k',t), \\
    \label{eq:PEFlux}
    \Pi_{\theta}(k,t) &= \sum_{k' \leq k} T_\theta(k',t),
\end{align}
\end{subequations}
where $T_u(k,t)$ and $T_\theta(k,t)$ are the 
non-linear kinetic and potential energy transfer across $k$: 
\begin{subequations}
\label{eq:Transfer}
\begin{align}
    \label{eq:KETransfer}
    T_u(k,t) &= -\sum_{k \leq |{\bf k}| < k+\Delta k} \widehat{\psi}^*_{\bf k}(t) \widehat{\{\psi,\nabla^2 \psi \}}_{\bf k}(t), \\
    \label{eq:PETransfer}
    T_\theta(k,t) &= \sum_{k \leq |{\bf k}| < k+\Delta k} \widehat{\theta}^*_{\bf k}(t) \widehat{\{\psi,\theta \}}_{\bf k}(t).
\end{align}
\end{subequations}
The notation $\widehat{\{ . \}}_{\bf k}$ represents the Fourier mode of the Poisson bracket expanded 
using Eq.~\eqref{eq:Basis}, and the asterisk denotes complex conjugation.

The spectra of the small-scale viscous dissipation $D_\nu(k,t)$, the large-scale friction $D_\mu(k,t)$ and the thermal dissipation $D_\kappa(k,t)$ are defined as
\begin{subequations}
\label{eq:FrictionTerms}
\begin{align}
    D_\nu(k,t) &= 2 \nu k^{2n}\what E_u(k,t), \\
    D_\mu(k,t) &= 2 \mu k^{-2} \what E_u(k,t), \\
    D_\kappa(k,t) &= 2 \kappa k^{2n} \what E_\theta(k,t),
\end{align}
\end{subequations}
and the buoyancy term $F_B$ is given by
\begin{align}
    \label{eq:BouyancyTerm}
    F_B(k,t) &= \sum_{k \leq |{\bf k}| < k+\Delta k} i k_x \widehat{\psi^*}_{\bf k}(t) \widehat{\theta}_{\bf k}(t). 
\end{align}

The Nusselt number is a dimensionless measure of the averaged vertical heat flux, defined mathematically by
\begin{equation}
    \Nu = 1 + \frac{\avg{\pd_x \psi \, \theta}}{\kappa \Delta T / L^{2n-1}}.
    \label{eq:Nusselt}
\end{equation}
Using the above definition, one can derive the following exact relations for the kinetic and potential energy balances in the statistically stationary regime, as in \cite{shraimansiggia1990}:
\begin{subequations}\label{eq:energy_eqns}
\begin{align}\label{eq:KEexact}
    \epsilon_u &= \epsilon_\nu + \epsilon_\mu
        = \frac{\nu^3}{L^{6n-2}}(\Nu - 1)\frac{\Ra}{\Pr^2}
    \\
    \label{eq:PEexact}
    \epsilon_\theta &= \epsilon_\kappa
    = \frac{\kappa \Delta T^2}{L^{2n}} (\Nu - 1)
\end{align}
\end{subequations}
where $\epsilon_u = \alpha g \avg{\psi \, \pd_x \theta}$ is the injection rate of kinetic energy due to buoyancy, $\epsilon_\nu = \nu {\langle\psi \nabla^{2(n+1)} \psi\rangle}$ is the viscous dissipation rate, $\epsilon_\mu = \mu {\langle\psi^2\rangle}$ is the large scale dissipation rate, $\epsilon_\theta = \frac{\Delta T}{L} \avg{\pd_x \psi \,\theta}$ is the injection rate of potential energy due to buoyancy and $\epsilon_\kappa = \kappa {\langle\theta \nabla^{2n} \theta\rangle}$ is the thermal dissipation rate.

\subsection{Elevator modes}

Upon linearising \eqref{eq:gov_eq} about the conductive state ($\psi=\theta=0$), we find that infinitesimal solutions with
$\psi(\mathbf{x},t)=e^{i \mathbf{k}\cdot\mathbf{x}+\sigma t}$
are possible provided the normalised linear growth rate $\sigma$ satisfies the relation
\begin{equation}\label{eq:disprel}
\left(\sigma+k^{2n}\right)
\left(\sigma+\Pr k^{2n}+\frac{\Rh\sqrt{\Ra\Pr}}{k^2}\right)
=\frac{\Ra\Pr (2\pi k_x)^2}{k^2},
\end{equation}
which has two real roots for $\sigma$. From these two roots, one is positive if and only if
\begin{equation}\label{eq:Rhineq1}
\Rh\sqrt{\frac{\Ra}{\Pr}}<
\frac{\Ra(2\pi k_x)^2}{k^{2n}}-k^{2n+2}.
\end{equation}
One can show that $\sigma$ is a monotonic decreasing function of $k_y$, so the most dangerous modes are independent of $y$, with
\begin{equation}
\psi(\mathbf{x},t)=e^{ik_x x+\sigma t},
\qquad k_x\in\mathbb{Z}_{>0}.
\end{equation}
Indeed, such a unidirectional mode satisfies the nonlinear governing equations \eqref{eq:gov_eq} \emph{exactly} (because the nonlinear Poisson bracket terms are identically zero).

Although the maximum growth rate does not necessarily occur at the minimum wavenumber $k_x=1$,
it is straightforward to show that \eqref{eq:Rhineq1} is satisfied for \emph{some} $(k_x,k_y)$ if and only if it is satisfied at $(k_x,k_y)=(1,0)$. In other words, exact solutions of the problem \eqref{eq:gov_eq} that grow exponentially without bound exist whenever
\begin{equation}\label{eq:Rhineq2}
\Rh\sqrt{\frac{\Ra}{\Pr}}<
\frac{\Ra}{(2\pi)^{2(n-1)}}-(2\pi)^{2n+2}.
\end{equation}

\section{Results}
\label{sec:results}

\subsection{Zero and infinite Prandtl number}
\label{sec:ZeroInfPr}

Before considering the $\Pr \to \infty$ limit, to simplify the analysis let's write
the non-dimensional form of Eqs. \eqref{eq:gov_eq} in accordance with Eqs. \eqref{eq:scaling}, which yield
\begin{subequations}
\label{eq:gov_eq_Spectra_ND}
\begin{align}
    \label{eq:gov_eq_Spectra_mom_ND}
    \pd_t \nabla^{2} \psi + \{\psi, \nabla^2 \psi\} &=  \Ra\Pr  \pd_x \theta + \Pr(-1)^{n+1}\nabla^{2(n+1)} \psi + \Rh \sqrt{\frac{\Ra}{\Pr}}\, \psi, \\
    \label{eq:gov_eq_Spectra_heat_ND}
    \pd_t \theta + \{\psi, \theta \} &=  \pd_x \psi + (-1)^{n+1}\nabla^{2n} \theta.
\end{align}
\end{subequations}
As we take $\Pr$ to infinity, we ensure that $\Rh \sqrt{\Ra/\Pr}$ is finite to maintain the effects of the large-scale dissipation. The system \eqref{eq:gov_eq_Spectra_ND} thus reduces to
\begin{subequations}
\label{eq:PrInf_gov}
\begin{align}
   \Ra \pd_x \theta  &= \left[(-1)^n \nabla^{2(n+1)}-\Rh \sqrt{\frac{\Ra}{\Pr}}\right]\psi, \\
    \pd_t \theta + \{\psi, \theta \} &=  \pd_x \psi + (-1)^{n+1}\nabla^{2n} \theta.
\end{align}
\end{subequations}

In the complementary limit of zero Prandtl number, we have to rescale the variables according to 
\begin{align}
    \{\psi,\theta,t \} \mapsto \{\Pr \,\psi,\Pr\,\theta,t/\Pr \}
\end{align}
before letting $\Pr \to 0$, which removes the time derivative and advective term from the heat transport equation \eqref{eq:gov_eq_Spectra_heat_ND}. As in the $\Pr \to \infty$ case, we maintain the effects of the large-scale dissipation by ensuring that $\Rh \sqrt{\Ra/\Pr}$ is finite. This process reduces the system \eqref{eq:gov_eq_Spectra_ND} to
\begin{subequations}
\label{eq:PrZero_gov}
\begin{align}
    \pd_t \nabla^{2} \psi + \{\psi, \nabla^2 \psi\} &=  \Ra \pd_x \theta + (-1)^{n+1} \nabla^{2(n+1)} \psi + \Rh \sqrt{\frac{\Ra}{\Pr}} \psi, \\
    \pd_x \psi &=  (-1)^{n}\nabla^{2n} \theta.
\end{align}
\end{subequations}

\begin{figure}
\centering
 \includegraphics[width=\linewidth]{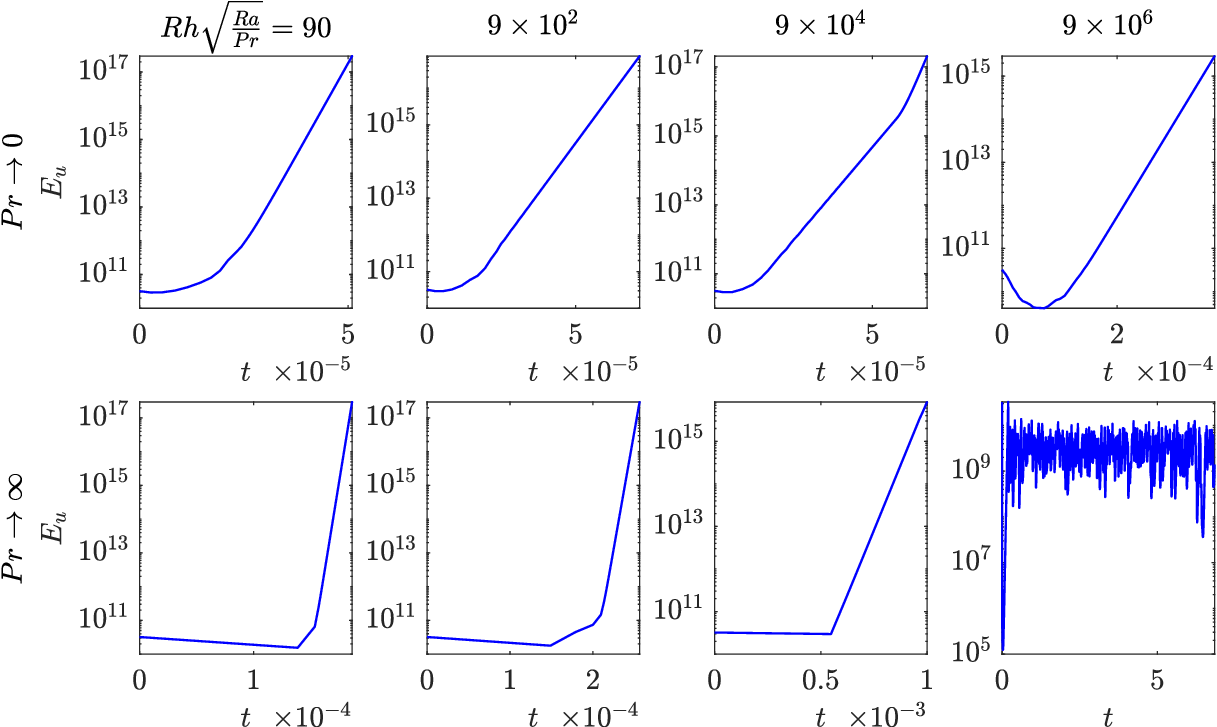}
 \caption[Time series at extreme Prandtl numbers]{
    Time series of the kinetic energy, $E_u$, in the $\Pr \to 0$ and $\Pr \to \infty$ limits with normal viscosity ($n=1$), at $\Ra =10^7$ and four different values of $\Rh \sqrt{\Ra/\Pr}$. The time series are initiated with random initial conditions.} 
 \label{fig:TSExremePr} 
\end{figure}

In Figure~\ref{fig:TSExremePr} we plot time series of the kinetic energy, $E_u$ in both the $\Pr \to 0$ and $\Pr \to \infty$ limits. We use normal viscosity (i.e. $n=1$), we set $\Ra =10^7$ for four different values of $\Rh \sqrt{\Ra/\Pr}$, and the simulations are initialised with random initial data.
The flow converges to a statistically steady state only in the case where
$\Pr \to \infty$ and $\Rh \sqrt{\Ra/\Pr} = 9 \times 10^6$, a value only $10\%$ smaller than the maximum value given by \eqref{eq:Rhineq2} at which all convection is suppressed. For all other parameter values attempted, an elevator mode takes over the dynamics and the energy grows without bound. 

At present, we are not able to reliably obtain turbulent saturated states in the extreme Prandtl number limits, unless the large-scale dissipation is made very strong. Instead, for the remainder of the paper we restrict our attention to finite values of Prandtl number, for which we find that $\Rh=(2\pi)^{5/2}$ is sufficient to prevent elevator modes and allow the system to reach turbulent saturated states.

\subsection{Finite Prandtl number: global variables \label{sec:global}}

\begin{figure}
\centering
\includegraphics[width=\linewidth]{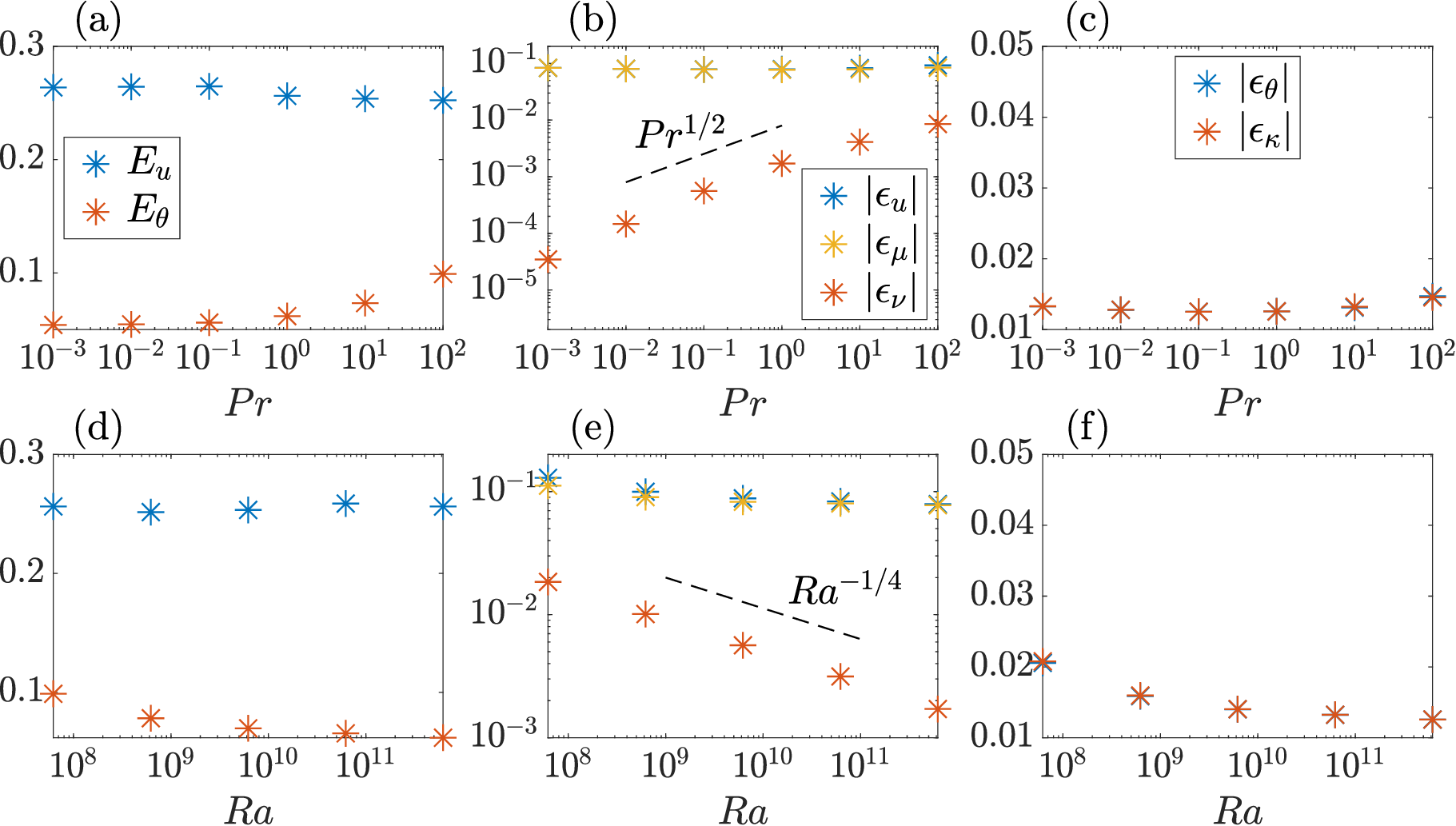}
 \caption{
Spatiotemporal averages
of the kinetic and potential energies and the terms in the kinetic and potential energy balances \eqref{eq:energy_eqns} as functions of $\Pr$ with $\Ra = 6.2 \times 10^{11}$ fixed (a--c) and as functions of $\Ra$ with $\Pr = 1$ fixed~(d--f). We use normal viscosity ($n=1$) and keep $\Rh  = (2\pi)^{5/2}$ fixed in all plots. }
 \label{fig:RaPrdependence}
\end{figure}

In Fig.~\ref{fig:RaPrdependence}%
(a)--(c) we show how the global quantities in the kinetic and potential energy balances \eqref{eq:energy_eqns} vary with $\Pr$ keeping $\Ra = 6.2 \times 10^{11}$,
while in (d)--(f) we show how the same quantities vary with $\Ra$ while keeping $\Pr=1$.
In both cases we use normal viscosity ($n=1$) and keep $\Rh=(2\pi)^{5/2}$ constant. 
Firstly, Fig.~\ref{fig:RaPrdependence}(a) and~(d) show that the kinetic and potential energies
remain virtually constant while $\Pr$ and $\Ra$ vary by at least four orders of magnitude. Secondly, Fig.~\ref{fig:RaPrdependence}(b) and~(e) show 
that $\epsilon_u \approx \epsilon_\mu$, which indicates that the majority of the kinetic energy, injected by buoyancy in the flow, is dissipated at large scales. This effect is caused by the inverse cascade of kinetic energy, which will be investigated in more detail in Sec.~\ref{sec:spectra}.
Note that $\epsilon_u$ and $\epsilon_\mu$ are almost independent of both $\Pr$ and $\Ra$, while the viscous dissipation scales like
\begin{equation}\label{eq:viscousdissipation}
    \epsilon_\nu \propto \Pr^{1/2} \Ra^{-1/4}.
\end{equation}
Finally, in Fig.~\ref{fig:RaPrdependence}(c) and~(f), we see that $\epsilon_\theta=\epsilon_\kappa$, as required by the potential energy balance \eqref{eq:PEexact}, and both quantities, 
are also virtually independent of both $\Pr$ and $\Ra$.

\begin{figure}
 \centering
 \includegraphics[width=0.7\linewidth]{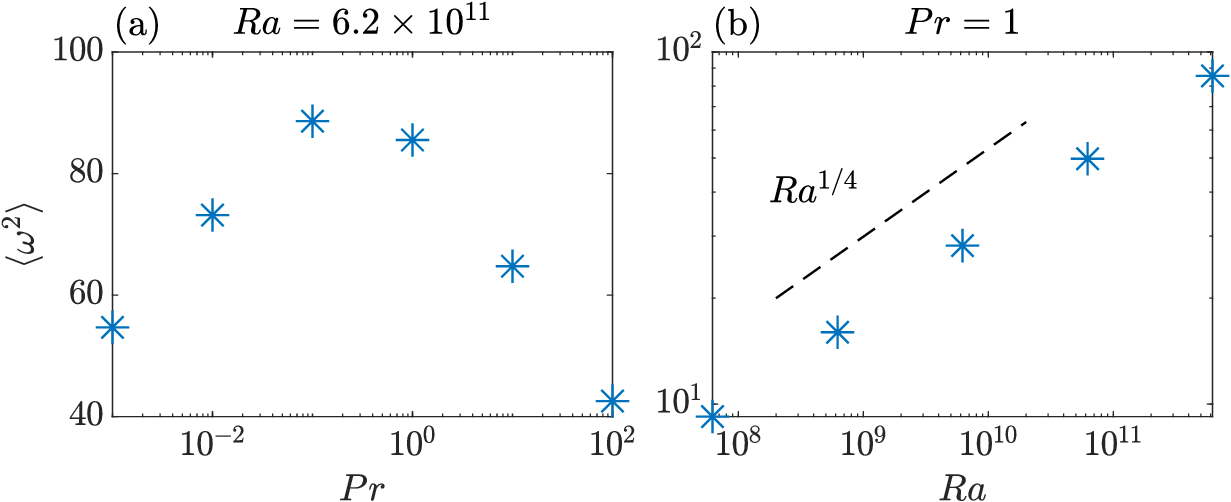}
 \caption{\label{fig:Enstrophy} 
Enstrophy $ \avg{\omega^2}$ as a function of $\Pr$ with $\Ra = 6.2 \times 10^{11}$ fixed (a), and as a function of $\Ra$ with $\Pr = 1$ fixed (b). We keep $n=1$ (normal viscosity) and $\Rh  = (2\pi)^{5/2}$ fixed in all plots.
}
\end{figure}

To explain the scaling of the viscous dissipation rate \eqref{eq:viscousdissipation},
we now look at how the enstrophy $\avg{\omega^2}$ varies with the Prandtl and Rayleigh numbers, where $\omega = \nabla^2 \psi$ is the vorticity of the flow. In Fig.~\ref{fig:Enstrophy}(a) we plot enstrophy 
as a function of $\Pr$ with $\Ra = 6.2 \times 10^{11}$ fixed, and in Fig.~\ref{fig:Enstrophy}(b)
as a function of $\Ra$ with $\Pr = 1$ fixed.
We keep $\Rh  = (2\pi)^{5/2}$ fixed throughout. From Fig.~\ref{fig:Enstrophy}(a) we observe that
enstrophy can be considered approximately independent of $\Pr$, because it varies by less than a factor of two over the five decades of Prandtl numbers considered, while Fig.~\ref{fig:Enstrophy}(b) demonstrates that enstrophy scales like $\Ra^{1/4}$, i.e.,
\begin{equation}\label{eq:enstrophy}
    \avg{\omega^2} \propto 
    \Pr^0 \Ra^{1/4}.
\end{equation}
With normal viscosity, by definition we have $\epsilon_\nu = \nu \avg{\psi \nabla^4 \psi} = \nu \avg{\omega^2}$ and, writing $\nu 
\propto (\Pr/\Ra)^{1/2}$,
we thus obtain the scaling of
Eq.~\eqref{eq:viscousdissipation} that is observed in Fig.~\ref{fig:RaPrdependence}.

\begin{figure}
\centering
 \includegraphics[width=0.7\linewidth]{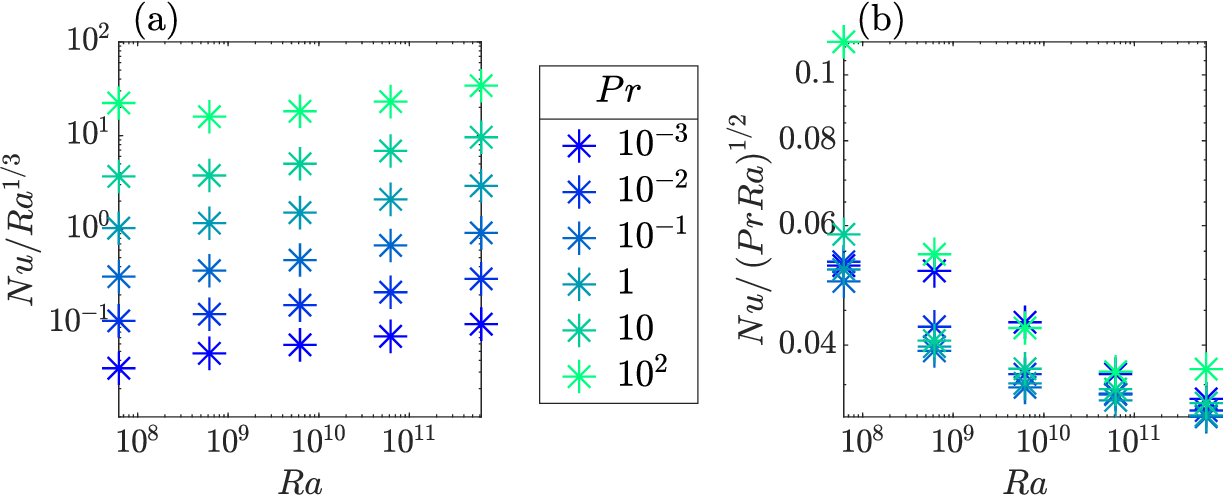}
 \caption{The Nusselt number compensated with the ‘classical’ and the ‘ultimate’ scaling in (a) and (b), respectively. The Prandtl numbers are colour-coded in the legend. }
  \label{fig:Nusselt}
\end{figure}

According to Fig.~\ref{fig:RaPrdependence}, the normalised energy injection rates $\epsilon_u$ and $\epsilon_\theta$ are both approximately constant in the range of $\Pr$ and $\Ra$ we considered.
With $\Ra\Pr \gg 1$ and $n = 1$, the net energy balances \eqref{eq:energy_eqns} thus produce the Nusselt number scaling
\begin{align}\label{eq:ultimate_scaling}
    \Nu \propto \Pr^{1/2}\Ra^{1/2}.
\end{align}
This relation is in agreement with the ultimate scaling. In Fig.~\ref{fig:Nusselt}, we plot the Nusselt number compensated by the classical scaling, $\Nu \propto \Pr^{0}\Ra^{1/3}$, and by the ultimate scaling, $\Nu \propto \Pr^{1/2}\Ra^{1/2}$. The ultimate scaling provides a much more convincing collapse of the data, with the fit becoming increasingly accurate as $\Ra$ increases.
Indeed, the ultimate scaling in terms of Rayleigh number dependence was expected to be exhibited by our simulations as we have effectively removed the boundary layers by applying periodic boundary conditions on the computational domain \citep{LohseToschi03,calzavarinietal05,schmidtetal12}. In addition, we demonstrate that the Prandtl number dependence follows the ultimate scaling, too.

\subsection{Finite Prandtl number: spectra \label{sec:spectra}}

\begin{figure}
 \centering
 \includegraphics[width=0.7\linewidth]{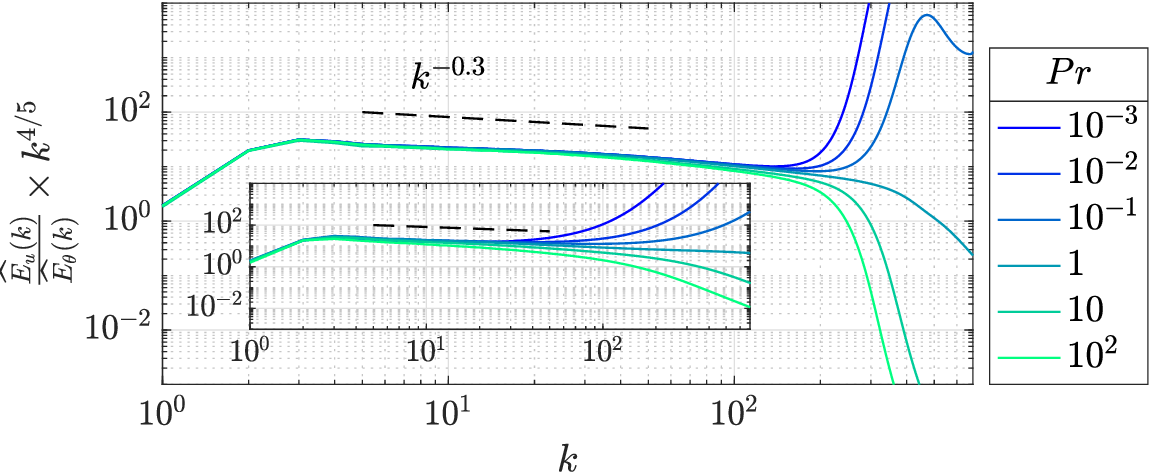}
 \caption{\label{fig:Ratio_Spectra} The ratio of the kinetic energy to the potential energy spectra $\what E_u(k)/\what E_\theta(k)$ compensated by $k^{4/5}$ for the hyperviscous simulations with $\Ra = 9.4 \times 10^{49}$, $\Rh  = (2\pi)^{5/2}$ and different Prandtl numbers as colour-coded in the legend. The inset shows the same ratio but for runs with normal viscosity at $\Ra = 6.2 \times 10^{11}$.
 }
\end{figure}

In this section, we examine the time-averaged kinetic and potential energy spectra and spectral fluxes. 
To eliminate finite Rayleigh number effects and to have large enough scale separation, we focus on the hyperviscous simulations (i.e., $n=4$ in Eqs.~\eqref{eq:gov_eq}) at $\Ra = 9.4 \times 10^{49}$ and $\Rh  = (2\pi)^{5/2}$. Results with normal viscosity  (i.e. $n=1$ in Eqs.~\eqref{eq:gov_eq}) at $\Ra = 6.2 \times 10^{11}$ are also presented for comparison.
According to the Bolgiano--Obukhov (BO) scaling \citep{Bolgiano59,Obukhov59}, the ratio of the kinetic to the potential energy spectra scales as
$\what E_u(k) / \what E_\theta(k) \propto
k^{-4/5}$, since
$\what E_u \propto
k^{-11/5}$ and
$\what E_\theta \propto
k^{-7/5}$. 
In Fig.~\ref{fig:Ratio_Spectra} we plot $\what E_u(k) / \what E_\theta(k)$ compensated by $k^{4/5}$ for the hyperviscous simulations and, in the inset, for the runs with normal viscosity. Instead of finding a wavenumber range where this scaling is valid, we observe a $k^{-0.3}$ power-law in the inertial range for all Prandtl numbers considered, leading us to the conclusion that BO scaling is not followed in our simulations. For the normal viscosity simulations we find the $k^{-0.3}$ power-law again to be followed, but within a narrower wavenumber range (see inset of Fig.~\ref{fig:Ratio_Spectra}). 
Fig.~\ref{fig:Ratio_Spectra} also shows that, for $\Pr \ll 1$, the kinetic energy is much larger than the potential energy at large wavenumbers. This is expected as the small scales are dominated by thermal diffusivity when $\Pr \ll 1$ and so the potential energy is dissipated much more effectively than the kinetic energy. The opposite is true for $\Pr \gg 1$.

\begin{figure}
 \centering
 \includegraphics[width=\linewidth]{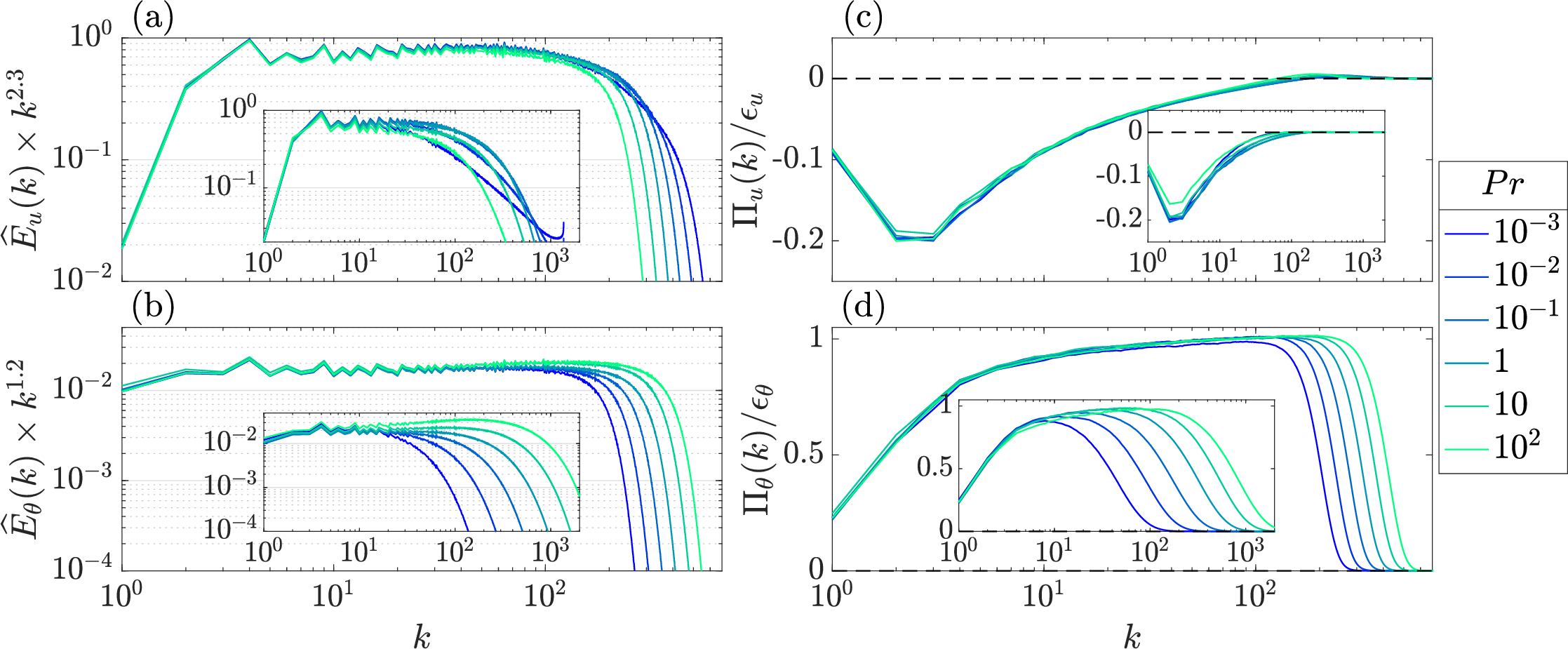}
 \caption{\label{fig:Flux_Spectra} The time averaged energy spectra compensated by best fit power laws (a,b), and spectral fluxes normalised by the time-averaged dissipation rates (c,d) for the runs with hyperviscosity at $\Ra = 9.4 \times 10^{49}$, $\Rh  = (2\pi)^{5/2}$ and different Prandtl numbers as colour-coded in the legend.
 The insets shows the same quantities but for runs with normal viscosity at $\Ra = 6.2 \times 10^{11}$.}
\end{figure}

In Fig.~\ref{fig:Flux_Spectra}(a) and~(b), we plot the kinetic and potential energy spectra multiplied by 
powers of $k$ chosen to best compensate for the power-laws exhibited by the spectra.
The hyperviscous runs are shown in the main plots, and normal viscosity runs in the insets. 
The observed behaviour, with $\hat{E}_u(k)\propto k^{-2.3}$ and $\hat{E}_\theta(k)\propto k^{-1.2}$,
is close to, but not fully consistent with, BO phenomenology, according to which the exponents should be $-11/5=-2.2$ and $-7/5=-1.4$, respectively. Moreover, the spectra we observe are in contrast to 3D RBC with periodic boundary conditions, where the kinetic and potential energy exhibit $k^{-5/3}$ spectra \citep{BorueOrszag97}, similar to those observed in passive scalar turbulence \citep{Warhaft00}.

To understand the turbulent cascades, in Fig.~\ref{fig:Flux_Spectra}(c) and (d), we plot the associated kinetic and potential energy fluxes normalised by the time-averaged injection rates of energy due to buoyancy. The positive potential energy flux suggests a strong direct cascade, while there is a weak inverse cascade of kinetic energy, which is typical for 2D turbulence \citep{Alexakis18}. For $\Pr = 1$ these types of cascades are in agreement with \cite{xiehuang22}. The negative kinetic energy flux peaks at low wavenumbers, while the potential energy flux peaks at high wavenumbers. 
The inverse cascade of kinetic energy is not affected by the Prandtl number; however, the direct cascade of potential energy moves to higher wavenumbers along with the peak of $\Pi_\theta(k)$ as $\Pr$ increases. 
We emphasise the wavenumber dependence of kinetic and potential energy fluxes, with $\pd_k \Pi_u(k) > 0$ and $\pd_k \Pi_\theta(k) > 0$ in the inertial range of wavenumbers for all of the Prandtl numbers considered.

\begin{figure}
 \centering
 \includegraphics[width=\linewidth]{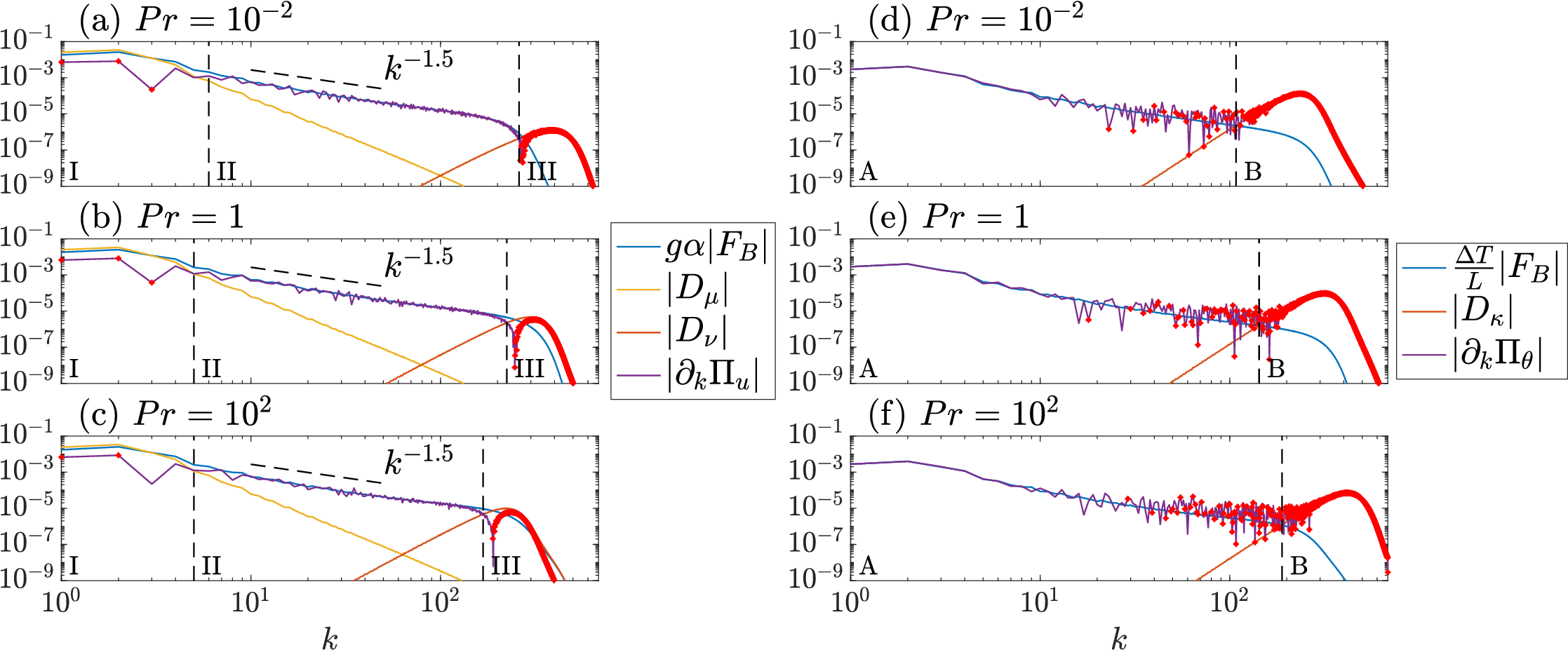}
 \caption{\label{fig:HV_Scales} 
Time-averaged spectra of the terms in the kinetic energy balance \eqref{eq:KESpecEvolution} (a)--(c); and the potential energy balance \eqref{eq:PESpecEvolution} (d)--(f) for the runs with hyperviscosity, \tog{\Ra = 9.4 \times 10^{49}},  $\Rh  = (2\pi)^{5/2}$, and $\Pr = 10^{-2}$ (a), (d), $\Pr = 1$ (b), (e), and $\Pr = 10^{2}$ (c), (f). The terms displayed in the legends are defined in \eqref{eq:Flux}, \eqref{eq:FrictionTerms} and \eqref{eq:BouyancyTerm}. In (a)--(c) and (d)--(f), we observe three and two distinct dominant balances, respectively, annotated I--III and A--B. The red dots indicate where the inertial terms become negative.
}
\end{figure}

\begin{figure}
 \centering
 \includegraphics[width=\linewidth]{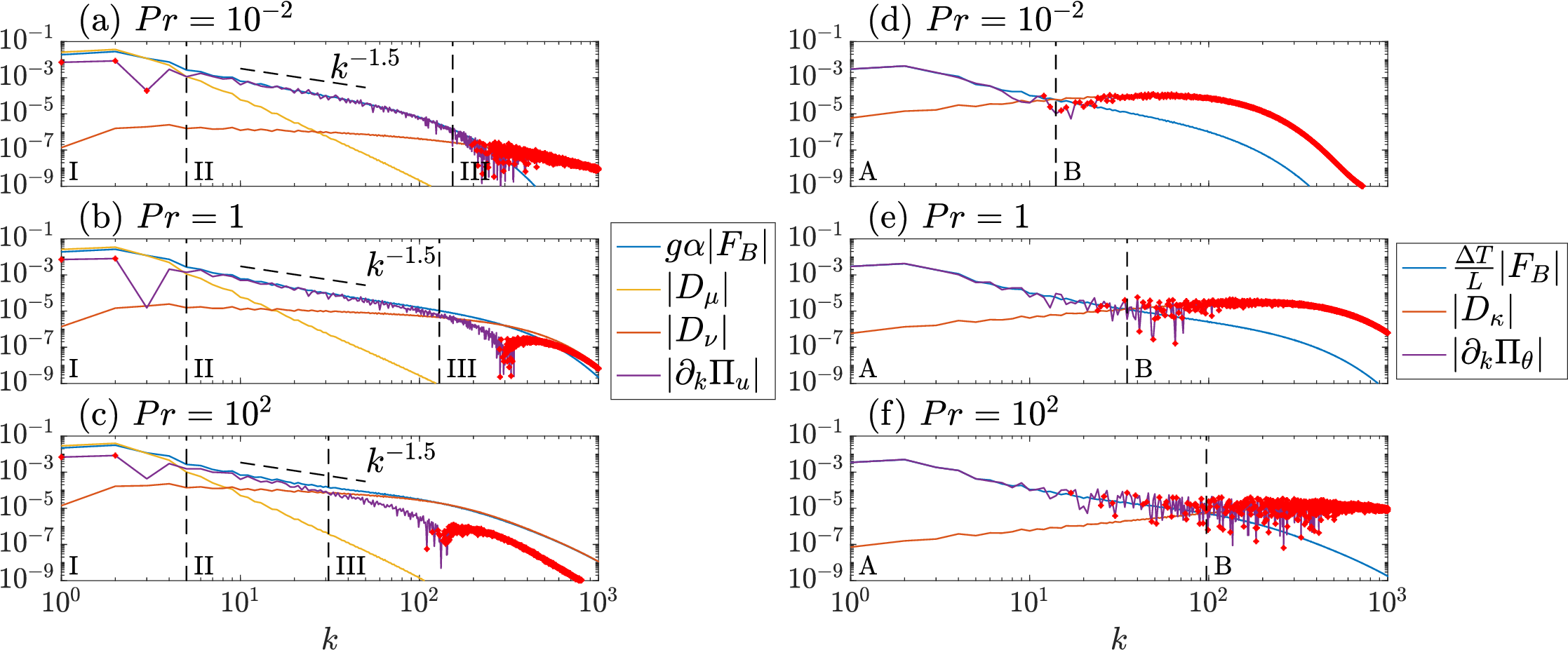}
 \caption{\label{fig:NV_Scales.eps}
 Time-averaged spectra of the terms in the kinetic energy balance \eqref{eq:KESpecEvolution} (a)--(c); and the potential energy balance \eqref{eq:PESpecEvolution} (d)--(f) for the runs with normal viscosity, \tog{\Ra = 6.2 \times 10^{11}},  $\Rh  = (2\pi)^{5/2}$, and $\Pr = 10^{-2}$ (a), (d), $\Pr = 1$ (b), (e), and $\Pr = 10^{2}$ (c), (f).
The terms displayed in the legend are defined in \eqref{eq:Flux}, \eqref{eq:FrictionTerms} and \eqref{eq:BouyancyTerm}. In (a)--(c) and (d)--(f), we observe three and two distinct dominant balances, respectively, annotated I--III and A--B. The red dots indicate where the inertial terms become negative.}
\end{figure}

In Fig.~\ref{fig:HV_Scales}, we present the time-averaged spectra of the magnitudes of the terms in the kinetic and potential energy balances \eqref{eq:SpecEvolution} for runs with hyperviscosity and three different values of
$\Pr\in\left\{10^{-2},1,10^2\right\}$.
The corresponding plots with normal viscosity and $\Ra=6.2\times10^{11}$ are shown in Fig.~\ref{fig:NV_Scales.eps}.
In both figures, the red dots indicate where the inertial flux terms $\partial_k\Pi_u$ and $\partial_k\Pi_\theta$ become negative.
In the kinetic energy balance, we identify three distinct wavenumber ranges, labeled I to III in the plots. 
In region I, the kinetic energy injected by buoyancy is dissipated by the large-scale friction.
In region II, the inertial term balances buoyancy, which is positive for all wavenumbers in this region, i.e.
\begin{align}
    \partial_k \Pi_u(k) \approx \alpha g F_B(k) > 0.
    \label{eq:utransplus}
\end{align}
This relation shows how the kinetic energy injected by buoyancy is cascaded to larger scales in the inertial range of wavenumbers and explains the $k$ dependence of the kinetic energy flux we see in Fig.~\ref{fig:Flux_Spectra}(c).
Note that region II is largest for small Prandtl numbers, especially in the runs with normal viscosity, as evident from the results presented in Fig.~\ref{fig:NV_Scales.eps}(a)--(c).
In region III, the balances between terms depend on the Prandtl number. 
For $\Pr = 10^{-2}$ buoyancy decays rapidly and so small-scale viscous dissipation is balanced by the inertial term, which is negative in this range of wavenumbers. As $\Pr$ increases, buoyancy becomes more significant in the balance of region III between the small-scale viscous dissipation and the inertial term. For $\Pr\gtrsim10^2$, small-scale viscous dissipation seems to be balanced by buoyancy rather than by the inertial term. This effect is shown more clearly in the runs with normal viscosity shown in Fig.~\ref{fig:NV_Scales.eps}(c),
where the small-scale dissipation range is much larger.

In the potential energy balance, we identify two distinct wavenumber ranges, labeled A and B in the plots (see (d)--(f) in Plots.~\ref{fig:HV_Scales} and~\ref{fig:NV_Scales.eps}). 
In these plots we observe the inertial term to be balanced by buoyancy in region A and by small-
scale thermal dissipation in region B.
In other words, the potential energy injected by buoyancy in region~A is cascaded to larger wavenumbers, where it is dissipated by thermal diffusivity.

We recall that the red dots in Plots.~\ref{fig:HV_Scales} and~\ref{fig:NV_Scales.eps} show where the inertial terms of the kinetic and potential energy become negative. This sign change occurs primarily
in region III for $\pd_k \Pi_u(k)$ and region B for $\pd_k \Pi_\theta(k)$, the latter corresponding to
the large negative gradient of $\Pi_\theta(k)$ observed in Fig.~\ref{fig:Flux_Spectra} at large wavenumbers.
Near the boundary between regions~A and~B in
Plots.~\ref{fig:HV_Scales} and~\ref{fig:NV_Scales.eps}(d)--(f), we observe that  $\pd_k \Pi_\theta(k)$
exhibits fluctuations between positive and negative values over these wavenumbers. However, for the majority of the wavenumbers in region A we find that
\begin{align}
    \partial_k \Pi_\theta(k) \approx \frac{\Delta T}{L} F_B(k) > 0.
    \label{eq:thetatransplus}
\end{align}
This relation explains the $k$ dependence of the potential energy flux we observe in Fig.~\ref{fig:Flux_Spectra}(d) and demonstrates that the BO phenomenology, which assumes that the potential energy flux is constant in the inertial range of wavenumbers, does not hold for any of the Prandtl numbers we studied.
Note that in the wavenumber range where region II and A overlap, we have $\partial_k \Pi_u(k)/\alpha g \approx L\partial_k \Pi_\theta(k) / \Delta T \approx F_B(k)$, such that $\Pi_u(k) - \frac{\alpha g L}{\Delta T} \Pi_\theta(k)$ is approximately constant for all Prandtl numbers considered. This is the inertial range of scales, where the viscous, diffusive and friction effects can be neglected and so the energy flux is constant. This is expected for the quantity $E_u - \frac{agL}{\Delta T} E_\theta$, which is constant in the limit of $\nu \to 0$, $\kappa \to 0$, and $\mu \to 0$.

\section{Conclusions \label{sec:Conclusions}}

In this paper we study the effects of varying the Prandtl and Rayleigh numbers on the dynamics of two-dimensional Rayleigh-B\'enard convection without boundaries, i.e., with periodic boundary conditions.
First, we focus on the limits of $\Pr \to 0$ and $\Pr \to \infty$. Our findings indicate that, unless large-scale dissipation is made so strong as to almost suppress convection completely, large-scale elevator modes dominate the dynamics. Such elevator modes 
have long been known \citep{BainesGill69}. 
In all parameter values simulated, the inequality \eqref{eq:Rhineq2} is violated, implying that the system admits exact single-mode solutions that grow exponentially without bound. Instead, we turn to finite Prandtl numbers, where 
we find that non-linear interactions between modes continue to allow the system to find a turbulent stationary state. In general, whether or not the solution blows up must depend on the initial conditions, in a way that is not currently understood.

Examining the Prandtl and Rayleigh number dependence of the terms in the kinetic and potential energy balances, we find that
the enstrophy scales as $\avg{\omega^2} \propto \Pr^0 \Ra^{1/4}$ and hence 
the small-scale viscous dissipation
scales as $\epsilon_\nu= \nu \avg{\omega^2} \propto \Pr^{1/2}\Ra^{-1/4}$.
On the other hand, we observe that the injection rate of kinetic energy $\epsilon_u$ due to buoyancy is effectively independent of both the Prandtl and the Rayleigh number. Using this observation,
we find that $\Nu \propto \Pr^{1/2}\Ra^{1/2}$, which agrees with the so-called ultimate scaling.

Looking at the kinetic and potential energy spectral fluxes, we find an inverse cascade of kinetic energy and a direct cascade of potential energy in contrast to 3D RBC \citep{VermaBook18}, where both kinetic and potential energies cascade toward small scales. The inverse cascade is independent of the Prandtl number, while the peak of the potential energy flux moves to higher wavenumbers as $\Pr$ increases. The kinetic and potential energy fluxes, $\Pi_u(k)$ and $\Pi_\theta(k)$, are not constant in the inertial range because both are balanced by the buoyancy  term $F_B(k)$,
which is predominantly positive in this range of wavenumbers. These two balances imply a positive slope in $k$ for the fluxes. Although we observe no range of wavenumbers where either $\Pi_u(k)$ or $\Pi_\theta(k)$ is constant, we find that they are connected by the relation $\Pi_u(k) - \frac{\alpha g  L}{\Delta T} \Pi_\theta(k) \approx \text{constant}$ in the overlap between regions~II and~A shown in Plots.~\ref{fig:HV_Scales} and~\ref{fig:NV_Scales.eps}.

The kinetic energy spectra scale as $\what E_u(k) \propto k^{-2.3}$, which is close to $k^{-11/5}$ behaviour of the BO phenomenology. However, the potential energy spectra scale as $\what E_\theta(k) \propto k^{-1.2}$, which deviates significantly from the $k^{-7/5}$ scaling predicted by the BO arguments. The deviation from the BO phenomenology is clearer when we test the scaling $\what E_u(k) / \what E_\theta(k) k^{4/5} \approx \text{constant}$, which is clearly not followed by our spectra which follow $\what E_u(k) / \what E_\theta(k) k^{4/5} \propto k^{-0.3}$. For the hyperviscous simulations,
the observed power-laws in the inertial range of the kinetic and potential energy spectra do not show any dependence on the Prandtl number. The only dependence we observe is at the dissipative range of wavenumbers, where the viscosity and the thermal diffusivity dominate the dynamics. In the spectra from the normal viscosity simulations, the effects of the Prandtl number are more significant due to the comparatively low scale separation. Hence, the inertial range over which a power-law behaviour can be observed is truncated. 

This study clearly demonstrates the necessity for large scale separation to be able to make clearer conclusions on the spectral dynamics and the power-law exponents of two-dimensional Rayleigh-B\'enard convection. This requirement makes similar studies in three dimensions more challenging. The development of a phenomenology where buoyancy acts as a broadband spectral forcing is required to interpret the current observations. Numerical simulations at Prandtl and Rayleigh numbers outside the ranges we investigated, i.e., $\Pr < 10^{-3}$, $\Pr > 10^2$ and $\Ra > 10^{12}$, are challenging but would be of great interest to see if they agree with the hyperviscous simulations we have performed and to provide a more complete picture of the asymptotic regime of buoyancy driven turbulent convection.

\bibliographystyle{jfm}
\bibliography{references}

\end{document}